# Improved electro-grafting of nitropyrene onto onion-like carbon via in situ electrochemical reduction and polymerization: Tailoring redox energy density of the supercapacitor positive electrode

Bihag Anothumakkool[a], Pierre-Louis Taberna[b], Barbara Daffos[b], Patrice Simon[b], Yuman Sayed-Ahmad-Baraza[a], Chris Ewels[a], Thierry Brousse*[a] and Joel Gaubicher[a]*

Herein, we report a improved method for the physical grafting of 1-nitropyrene (Pyr-NO$_2$) onto highly graphitized carbon onion. This is achieved through a lowering of the onset potential of the pyrene polymerization via in situ reduction of the NO$_2$ group. The additional redox activity pertaining to the reduced NO$_2$ enables exceeding the faradaic capacity which is associated with the p-doping of the grafted pyrene backbone, as observed for pyrene, 1-aminopyrene, and unreduced Pyr-NO$_2$. Theoretical calculations demonstrate the charge transfer and binding enthalpy capabilities of Pyr-NO$_2$, which are significantly higher than those of the other two species, and which allow for improved p-stacking on the carbon surface. Upon 20 wt % grafting of Pyr-NO$_2$, the capacity of the electrode jumps from 20 mAh g$^{-1}_{electrode}$ to 38 mAh g$^{-1}_{electrode}$, which corresponds to 110 mAh g$^{-1}$ per mass of Pyr-NO$_2$ and the average potential is increased by 200 mV. Very interestingly, this high performance is also coupled with outstanding retention with respect to both the initial capacity for more than 4000 cycles, as well as the power characteristics, demonstrating the considerable advantages of employing the present in situ grafting technique.

Highly efficient electrochemical energy storage devices are an integral part of renewable energy technology, as they smooth out its intermittent nature by effective storage and delivery[1,2]. Among such devices, supercapacitors[3] in comparison to Li-ion batteries show superior power density, excellent shelf life, high coulombic efficiency etc., which make them an ideal candidate for high-power applications. However, the low energy density (~5-8 Wh kg$^{-1}$) of conventional electrochemical double-layer capacitors (EDLC)[2,4] is a key issue with respect to hindering their use in a higher energy bracket. Molecular redox grafting[5-12] is one of the strategies currently used to enable higher energy performance, while the other consists in tailoring the micropore size[13]. Indeed, the faradaic processes associated with redox grafting convey both higher charge storage and a higher average voltage. In this regard, conducting polymers[14], as well as various metal-based oxide compounds[15], are considered to be potential alternative materials as many of them show relativity high redox capacity. However, most of these materials only show superior electrochemical properties in aqueous electrolytes, which severely restrict the potential window and, therefore, the final energy density of the supercapacitor. One of the main reasons for these additives not being particularly efficient in organic media (they only convey higher charge transfer capacity) pertains to the value of their redox potential, which lies in the middle of the electrochemical window of the electrolyte.

Thus, the selection of materials possessing more extreme redox potentials, along with high capacity and cycling stability, are essential criteria for achieving competitive performance. As is the case of (2,2,6,6-Tetramethylpiperidin-1-yl)oxyl (TEMPO), many organic centers show ideal redox activity in non-aqueous media[16] however, they presently require a polymer backbone, which shows low conductivity, and thereby restricts the use of these TEMPO derivatives to either low-power applications or low surface capacity thin-film electrodes[17]. Moreover, the final cost of the device and availability of the material often prove detrimental to widespread use, which is a major concern with respect to the above molecules as they require extensive synthetic chemistry.

Pyrene, one of the polycyclic aromatic compounds, is known for its stable redox chemistry in a polymerized state at an ideal potential, and thus for its propensity to serve as positive electrode[18]. Additionally, pyrene derivatives develop π-stacking type interactions, especially on graphitic carbon, and can be further polymerized to form a p-doped conducting polymer via simple electrochemical methods. Our group recently reported a strategy for *in situ* grafting (here onwards the term grafting indicate the non covalent π stacking of pyrene molecules) of such pyrene derivatives onto a carbon fiber electrode using a conventional electrolyte[6]. This approach offers enormous advantages with respect to upscaling considerations because it does not impose any change in electrode chemistry protocols, as opposed to those incurred by *ex situ* chemistries involving pyrene units, as reported by other researchers[19]. Considering that the adsorption [20] of polycyclic pyrene from the electrolyte is driven by π-π interactions, the graphitization state of the carbon substrate[21] is obviously a critical parameter. Among the various types of carbon[4] utilized for the EDLC, carbon onions/onion-like carbons(OLCs)[22] are known for exhibiting a pore size distribution in the mesoporosity range (2-12 nm), arising from the voids between the particles. They possess highly graphitized graphene layers arranged in spherical onion-like structures of 5–10 nm particles with a relatively high specific surface area (SSA, 640 m$^2$ g$^{-1}$). Furthermore, even though they show fairly low capacitance (~30 F g$^{-1}$)[23], they are widely recognized for their high power capability[24, 25], in contrast to activated carbons that possess an amorphous surface and a tortuous micropore network. Indeed, as inferred from the Bruggeman Model [26, 27], depending on the electrolyte conductivity, the high tortuosity of the carbon may be a reason for limited mass transport. Another distinct advantage of OLCs concerns their pack density (1.5-2.3 g cm$^{-3}$), which is much higher than that of graphene ~0.069 g cm$^{-3}$ [28], and even that of conventional activated porous carbon (0.5 to 0.7 g cm$^{-3}$). For these reasons, OLC appears to be an ideal material for the present redox molecular grafting strategy. Thus, taking these considerations in account, this article presents a method for doubling the capacity of a highly graphitic OLC sample, without degrading its power capabilities, by means of the controlled *in situ* grafting of various pyrene derivatives. During *in situ* reduction and polymerization, 1-nitropyrene is shown to play a pivotal role in developing high energy density positive electrodes for supercapacitors.

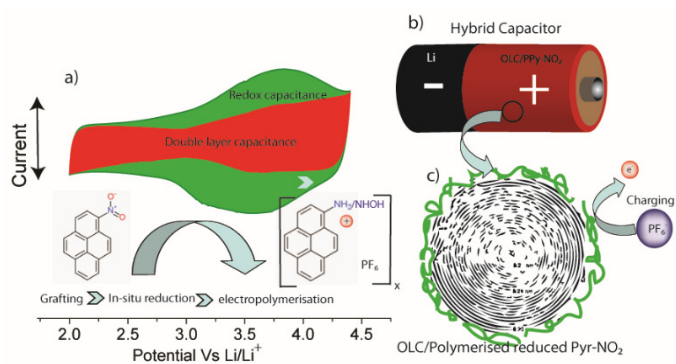

Scheme. a) Added redox capacity from 1-nitropyrene over EDLC of OLC represented in cyclic voltammetry profile; transformations of the same molecule in during cell cycling. b) Representation of the hybrid capacitor combination and b) schematics of OLC grafted with the polymerized pyrene.

**EXPERIMENTAL SECTION**

**Materials:** Pyrene, 1-aminopyrene, 1-nitropyrene were purchased from Aldrich Chemicals, LP-30 from BASF, and nanodiamond (>97 %) from PlasmaChem.

**Synthesis of carbon onion**: Commercial nanodiamond powder underwent a thermal treatment under vacuum in a graphite micro-furnace. A heating rate of 100°C min$^{-1}$ was used to attain a final temperature of 2000°C. The dwell time applied was 30 minutes.

**In situ electrografting**: A Bio-Logic VMP-2 or -3 was used for all of the electrochemical characterizations. Li metal was used as the reference and counter electrode. All the potentials mentioned hereafter were measured with either a Li/Li$^+$ reference electrode or with another reference electrode detailed in the corresponding paragraph. *In situ* grafting was carried out by electro-chemical polymerization in a 3-electrode Swagelok cell by cyclic voltammetry at a scan rate of 5 mV s$^{-1}$ in the potential range of 2 to 4.4 V (vs. Li). The electrolyte solution consisted of a mixture of pyrene (Pyr), 1-nitropyrene (Pyr-NO$_2$) or 1-aminopyerene (Pyr-NH$_2$) in LP-30. Electrochemical impedance (EIS) analysis was conducted from 100 KHz to 0.01 Hz against the open circuit potential with a sinus amplitude of 10 mV (Vrms = 7.07 mV). Charge-discharge cycling was performed between 1.5-4.4 V (vs. Li) with various current densities. The amount of grafted molecules was quantified using UV-visible spectroscopy (Perkin-Elmer Lambda 1050 with 3D WB detector module), by calibrating the characteristic peaks of pyrene in LP-30 electrolyte before and after *in situ* grafting. Scanning electron microscopy (SEM) and transmission electron microscopy (TEM) were carried out in Jeol JSM-7600F and Hitachi H9000NAR microscopes (300 kV, LaB$_6$, point to point resolution=0.18 nm), respectively.

**Density Functional Theory (DFT) Calculations of Pyrene and its derivatives:** DFT calculations were performed on large 128-atom hexagonal 8x8 graphene cells, with a 4x4x1 k-point grid, using a 0.04 eV Fermi smearing function for the electron temperature to aid self-consistent convergence. Interlayer spacing was set to over 12Å to avoid interaction with neighboring sheets. Charge density was constructed on a real-space grid with an energy cut-off of 175 Ha (200 Ha when oxygen present), while Kohn-Sham wave functions were constructed using localized Gaussian-based orbital functions (38/12/40/40 for C/H/O/N respectively, up to maximum angular momentum l=2). All atoms were fully optimized using a conjugate gradient algorithm. Atomic charge states were calculated using Mulliken population analysis.

**RESULT AND DISCUSSION**

Prior to electrochemical polymerization, all cells were kept in open-circuit potential (OCP) for 12 hours in order to ensure proper π-stacking of monomers onto the host OLC. Whenever the molecular grafting of a carbon surface is involved, a recurring issue arises relative to its impact on the electrical properties of the substrate. During OCV, it is instructive to trace the evolution of the charge transfer resistance (Rct) from EIS before and after the introduction of 1-nitropyrene (Pyr-NO$_2$) into the LP-30 electrolyte. As shown in **Figure 1a**, even in pure LP-30, there is a slight increase in the Rct, which could be tentatively ascribed to the interaction between solvated ions from the electrolyte and electrons in the vicinity of the carbon surface[29]. However, after the introduction of Pyr-NO$_2$, a significant increase in the charge transfer resistance (**Figure 1a**) is observed, indicating a possible electron trapping effect that could arise from polar Pyr-NO$_2$ molecules interacting with the OLC surface through π-stacking[30]. This molecular surface coverage can then serve to provide a partial definition of the potential of the working electrode from the redox potential of the molecule (Pyr-NO$_2$ around 2.0 V Vs Li/Li$^+$). Such surface grafting interactions are confirmed by the sharp 100 mV drop in OCV upon the addition of molecules to the medium (**Figure 1a**). As expected, such a drop was absent in case of softer carbon (YP-80), which demonstrates poor grafting, thus indicating that the OCV drop relates to the grafting of Pyr-NO$_2$ (Electronic Supporting Information; ESI† **Figure S1a**). To rule out the possibility of Li or electrolyte interactions with the Pyr-NO$_2$ being responsible for the drop in OCV, an experiment was also carried out in acetonitrile/TEABF$_4$ using an Ag/AgCl reference electrode and a stainless steel counter electrode (**Figure S1b**). A similar drop in OCV and increase in Rct was observed, which ruled out the possible role of both Li and electrolyte, and thereby confirmed the interaction of the OLC with Pyr-NO$_2$. The OCV subsequently stabilizes after approximately 90 minutes, indicating that the Pyr-NO$_2$ adsorption is fairly rapid at room temperature. In order to gain further insight into the adsorption kinetics, *in situ* UV-visible spectroscopy (**Figure S2**†) was conducted in the LP-30 for 14 hours. Although all molecules clearly show adsorption onto the graphitic OLC, the extent of that of the Pyr-NO$_2$ appears to be slightly higher. This higher affinity stems presumably from the high polarity of the Pyr-NO$_2$. A simple mixing of OLC with Pyr-NO$_2$ in LP-30 shows a 13 wt % uptake maximum over 7 days, which is much higher than that of Pyr (6.5 wt %) and Pyr-NH$_2$ (4.3 wt %). The number of moles of Pyr-NO$_2$ on the OLC surface is found to be 9.25 x 10$^{-5}$ moles cm$^{-2}$. This value indicates multilayer π stacking by considering the geometrical area of the molecules as 189 Å$^2$.

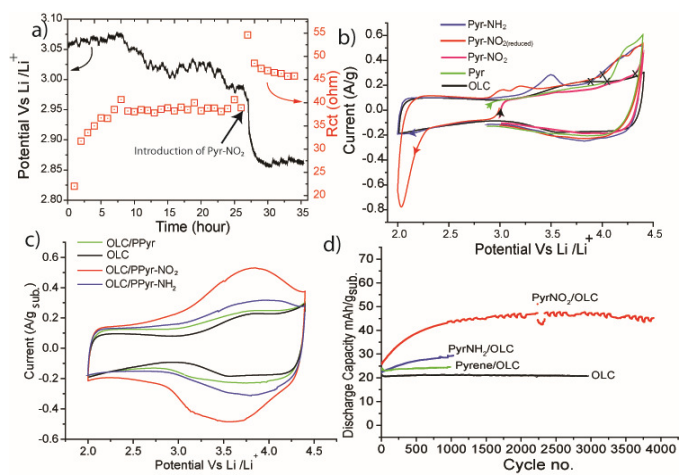

**Figure 1.** (a) Charge transfer resistance (calculated from EIS spectra) and open circuit potential of OLC electrode in LP-30 for 26 h, and further after the introduction of 2mM Pyr-NO$_2$ in LP-30. (b) 1$^{st}$ and (c) 1000$^{th}$ CV profile at a scan rate of 5 mV s$^{-1}$ in LP-30 containing 2mM pyrene or its derivates; the 'x' indicates the onset of polymerization. (d) The discharge capacity of OLC between 2-4.4 V during the continuous CV cycles carried out at 5 mV s$^{-1}$ in various pyrene samples in LP-30.

After the physical grafting, the electrodes underwent electrochemical polymerization in the same cell, i.e. an LP-30 containing 2 mM pyrene derivatives with Li metal as the counter and reference electrode. **Figure 1b** shows the initial CV profile at a scan rate of 5 mV s$^{-1}$ for a 5 mg cm$^{-2}$ electrode. Given that all the present pyrene derivatives have nearly the same surface (189, 212 and 200 Å$^2$ for Pyr, Pyr-NO$_2$ and Pyr-NH$_2$ respectively) available for radical stabilization through delocalization, the onset oxidation potential is expected to vary according to the Hammett constant, which relates to both the electron donating or withdrawing ability and resonance stabilization effects of the substituent [19]. The latter is 0.78 for –NO$_2$ and -0.66 for NH$_2$[31]. Accordingly, the onset potential for polymerization (*vs* Li/Li$^+$) was found to decrease in the following order: Pyr-NO$_2$ (4.33V) > Pyr (4.06V) > Pyr-NH$_2$ (3.86V). Upon oxidative electro-polymerization, the resulting polymer (oligomer) undergoes subsequent p-doping counterbalanced by the mass transport of PF$_6^-$ anions within its porosity, in the vicinity of 3.7 V (**Figure 1c**)[32]. However, it is worth noting that the corresponding faradaic capacity is directly superimposed onto that of the double layer, without any significant alteration to the latter. Another advantage of using pyrene derivatives lies in its versatility with respect to the choice of possible functionalization. In particular, redox units can still further enhance the faradaic capacity[6]. In the present work, we have investigated the role of the NO$_2$ group, which can undergo reduction below 2.4 V through various mechanisms to form either amino or nitro radical anion, or hydroxylamine, depending on the electrolytic medium[33]. In order to investigate the effect of NO$_2$ reduction on the electro-polymerization mechanism, a cell containing 2 mM Pyr-NO$_2$ in the electrolyte was biased in the negative direction, which indeed caused a reduction of the nitro group at a potential of around 2.3 V (**Figure 1b**). The product formed upon the first reduction of the nitro group is mainly hydroxylamine [34], which is calculated using the integrated capacity below the peak at 2.2V. This will further oxidized by losing 2e$^-$ at 3.2 V in the subsequent scan[35] (**Figure S3** in ESI†). The latter has a direct impact on the onset of the oxidative polymerization potential, which is lowered by approximately 33 mV in the preceding +ve scan. Such a change in the potential facilitates the polymerization current which is found to be nearly 4 times higher in case of reduced –NO$_2$ (12 C g$^{-1}$) compared to unreduced –NO$_2$ (3 C g$^{-1}$). However, the hydroxylamine is further reduced to amino during the ensuing cycles (the 2$^{nd}$ cycle is given **Figure S3** in ESI†). The proton associated with the formation of the hydroxylamine and amino group could possibly arise from the traces of water, the impurity, present in the LP-30 (20 ppm, which is 1.3 mM under our conditions). A more realistic assumption, however, would be to consider the effect of the protons released during oligomerization/polymerization of the pyrene unit (2 H$^+$ per pyrene would correspond to 8 μM under our conditions)[30]. Whichever the case may be, the nitro group certainly allows for an unprecedented improvement of the OLC capacity, and one which vastly exceeds that of Pyr-NH$_2$ and Pyr while cycling between 2-4.4V (**Figure 1d**). A comparative CV profile of the 1000$^{th}$ cycle is shown in **Figure 1c**.

In order to further clarify the interaction between the carbon surface and pyrene molecules, DFT calculations of Pyrene and its derivatives were performed. Given the large diameter of the carbon onions, local curvature effects on the external surface are negligible, and local molecular-surface interactions can be dealt with using graphene as a model for the surface. All three species (Pyr, Pyr-$NO_2$ and Pyr-$NH_2$) bind to the graphene and demonstrate charge transfer from the graphene to the molecules (see Table 1). In all cases, the most stable structure is one where the pyrene moiety is parallel to the graphene surface, thereby maximizing π-stacking interactions. The binding energy is largely independent of the precise molecular position (with a slight preference for AB stacking) suggesting that the molecules should be surface-mobile. Nonetheless, there are significant differences in their responses. Notably, both charge transfer and binding for Pyr-$NO_2$ is significantly higher than for the other two species (see Table 1), and which are consistent with experimental data obtained in the electrochemical and adsorption experiments (UV-visible studies). This is also reflected by its location, which is significantly closer to the graphene surface by comparison to the other two, and indeed results in slight distortion of the graphene basal plane, as can be seen in **Figure 2**. It should be noted that these calculations do not incorporate interactions with localized defects and impurities, which are likely to result in still higher binding energies. Additionally, if the molecules are able to migrate into positions lying between neighboring sheets (neighboring carbon onions in the experiment), one would expect a further increase in binding and associated charge transfer.

Table: 1: Summery of DFT calculations

|  | Charge transfer from graphene per molecule in electrons | Binding enthalpy between molecule and graphene surface (eV) | Average distance from moiety to surface (Angstroms) |
|---|---|---|---|
| Pyrene (Pyr) | 0.01 | 0.720 | 3.139 |
| 1-Nitropyrene (Pyr-$NO_2$) | 0.09 | 0.913 | 3.065 |
| 1-Aminopyrene (Pyr-$NH_2$) | 0.02 | 0.845 | 3.107 |

Post-mortem analysis of electrodes was carried out using SEM. **Figure 3a** shows an image of the bare OLC. After *in situ* polymerization of Pyr-$NO_2$ from LP-30 onto the carbon electrode (referred to as OLC/PPy-$NO_2$), the surface of the OLC is uniformly covered by the Pyr-$NO_2$, as is evident from **Figure 2b**. The wt % of polymer in this example is nearly 30 wt %, as inferred from UV-visible spectroscopy. Although the TEM observation of OLC (**Figure 3c**) allows for a clear depiction of the graphitic layers of the OLC before polymerization, the heavy polymer coverage prevents from a thorough characterization of the OLC/PPy-$NO_2$ surface. Further, XPS spectra unambiguously prove the reduction of $NO_2$ groups. The Nitrogen 1S spectrum of the polymerized electrode is shown in **Figure 4a** (N1S of pure Pyr-$NH_2$ and Pyr-$NO_2$ is shown in E**SI Figure S4**†). Binding energy (BE) of N1S is reduced from 406.3 eV (-$NO_2$) to 400.6 eV in the case of OLC/PPyr-$NO_2$, which can deconvoluted into two peaks issuing from -NHOH (402.2 eV) and –$NH_2$ (400.4 eV)[36]. The presence of a strong -$NH_2$ peak also confirms the conversion of NHOH formed upon the first cycle to -NH$_2$ during the remaining cycles. Raman spectra (**Figure 4b**) reconfirm the results obtained in the microscopy analysis. Spectra show typical D, G and 2D bands for the OLC as expected at wavelengths of 1335, 1571 and 2675 cm$^{-1}$, respectively[37]. Pure PPyr-NO$_2$, however, which was prepared for comparison by electropolymerization of Pyr-NO$_2$ on a Pt disc, shows enhanced background absorption due to the fluorescent effect, while two strong and broad peaks also appeared. A strong peak at 1377 cm$^{-1}$ corresponds to the symmetric stretching of N-O issuing from the NHOH[38]. Meanwhile, the second peak appeared to be a combination of two peaks at a frequency of 1620 and 1593 cm$^{-1}$, which corresponds to the C-C stretching of pyrene [38, 39]. In case of OLC/Pyr-NO$_2$, the background-enhanced spectrum is dominated by the polymer, though it is less prominent when compared to pure PPyr-NO$_2$. This confirms the nearly complete coverage of the carbon surface by PPyr-NO$_2$, thus impeding D and G signals from the inner carbon surface, and we speculate that the trace signals are probably merged into the strong peaks of the polymer. However, a trace of the 2D band confirms the presence of carbon onion in the sample. Interestingly, the peak at 1377 cm$^{-1}$ is found to be red shifted to 1348 cm$^{-1}$, while other bands are found to be unaffected and indicating the interaction of NO$_2$ with the carbon surface, as reported elsewhere[40].

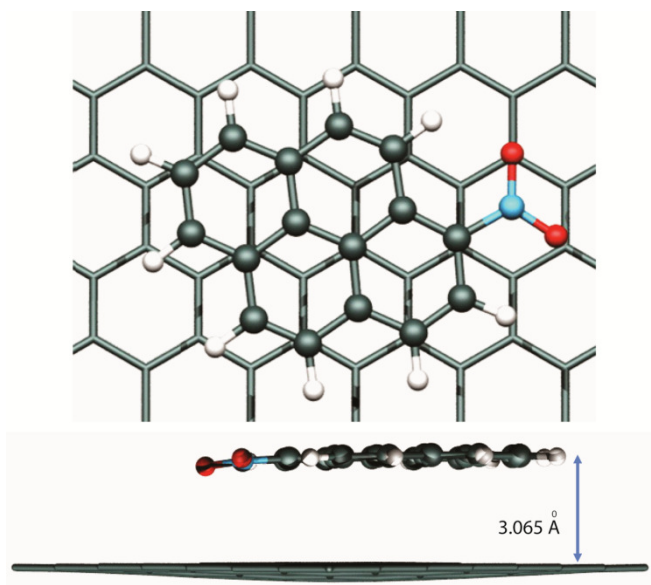

Figure 2. Stable AB-stacked orientation of 1-Nitropyrene on the graphene surface.

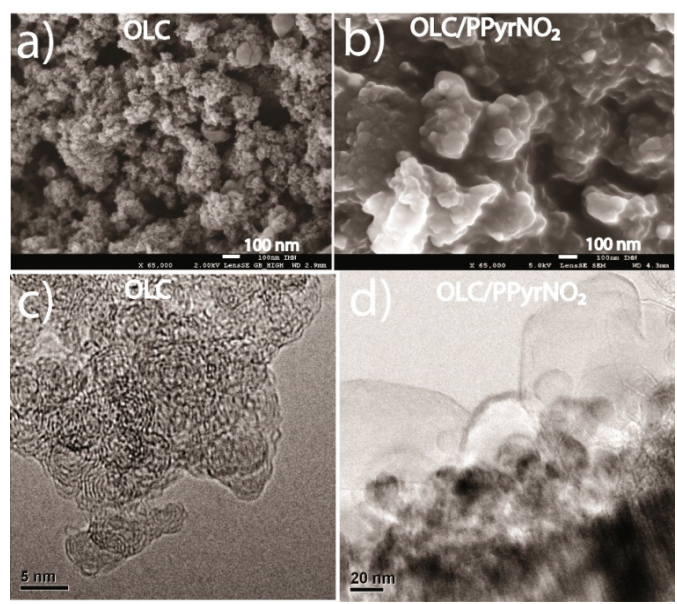

Figure 3. SEM images of: (a) bare OLC; (b) PPyr-NO2-coated OLC after electro-polymerization and TEM of (c) bare OLC; (d) PPyr-NO2-coated OLC after electro-polymerization.

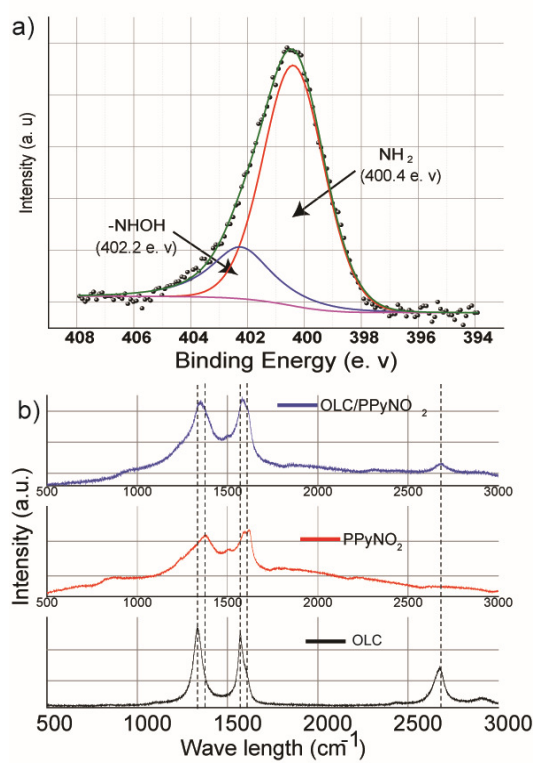

Figure 4. (a) N1S core XPS spectra of PPyr-NO2-coated OLC after electro-polymerization. (b) Raman spectra of OLC, electro-polymerized PyNO2 on Pt disc and OLC/PPy-NO2.

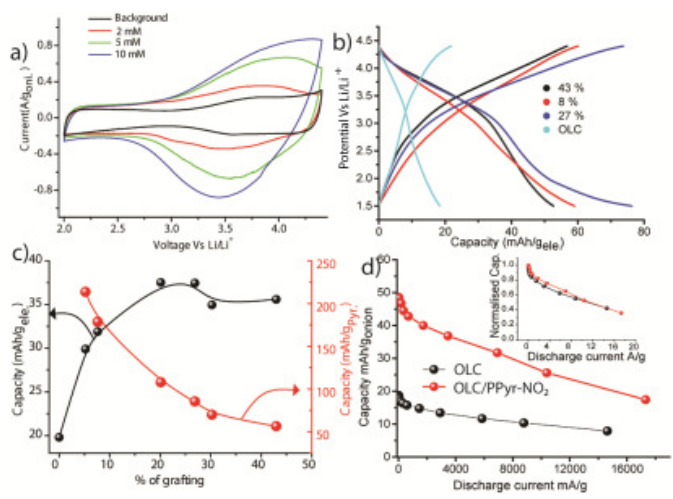

Figure 5. (a) 400$^{th}$ CV profile at a scan rate of 5 mV s-1 in LP-30 containing 2-10 mM Pyr-NO$_2$. (b) Charge-discharge plots at a current density of 0.1 mA cm$^{-2}$ of the OLC and its composites containing grafted PPy-NO$_2$, where the y-axis is normalized with the total mass of the electrode (OLC+PPyr-NO$_2$). (c) Capacity of the total electrode and PPy-NO$_2$ with varying amounts of grafting. (d) Capacity of the OLC and the 20 % grafted OLC with different current densities. The normalized capacity change is depicted in the inset.

The improvement of the overall capacity of OLC/PPy-NO$_2$ is found to reach a plateau after 1000 cycles, presumably because electron transfer through the polymer layer becomes increasingly hindered. Interestingly, extended cycling shows that the redox activity remains remarkably stable for more than 4000 cycles (**Figure 1d**). UV spectroscopy of the electrolyte shows a decreasing concentration of Py-NO$_2$ from 2 mM to 0.65 mM, indicating a coverage of 1.6 x 10$^{-10}$ mole cm$^{-2}$. In order to examine the range of possible grafting yield of the Pyr-NO$_2$ onto OLC, polymerization was carried out with a higher concentration of Pyr-NO2 in LP-30. A grafting of 27 and 43 wt % were measured in the case of 5 and 10 mM Pyr-NO$_2$ in LP-30 respectively, after 500 cycles of CV at a scan rate of 5 mV s$^{-1}$. The CV profiles, compared in Figure 5a, clearly illustrate the fact that a larger polarization appears above 20 wt % coverage. As regards the 20 wt % grafting, the capacity of the OLC electrode improves from 20 mAh g$^{-1}$_electrode to 47 mAh g$^{-1}$_electrode, (i.e. +90 %) (Figure 5b).

In order to confirm the superior role of graphitization over SSA towards the π-stacking and NO$_2$ reduction, a controlled experiment was carried out with highly porous carbon (YP-80 and microporous carbon from ACS chemicals). Even though porous carbon possesses SSA over 2000 m$^2$ g$^{-1}$, a negligible NO$_2$ reduction is observed due to poor π-stacking, and results in polymerization was carried out with a higher concentration of Pyr-NO$_2$ in LP-30 (**Figure S5†**). Thus, the higher capacity of the Pyr-NO$_2$ samples arises from three factors: (i) better π-stacking ability, as proved by the UV sorption studies; (ii) lower oxidation potential after reduction of NO$_2$ functionalities, both contributing to enhancing the graftingyield; and (iii) two redox reactions (associated with the reduced NO$_2$ and the pyrene polymer). In order to validate this theory, Pyr-NO$_2$ was cycled with a cut-off discharge potential of 3V, which thereby prevents the reduction of the nitro groups. As expected, the onset of the polymerization occurs at a higher potential (4.33 V), but more importantly, the capacity improvement is found to be greatly reduced compared to that corresponding to the 2V cut-off potential (SI **Figure S6†**). XPS analysis of the above electrode after 800 cycles indicate only a partial reduction of –$NO_2$ groups. A strong peak at 406 eV of N1S XPS in **Figure S7†** even after long cycles confirms this fact. As for the two other derivatives, the capacity of the OLC functionalized with Pyr-$NH_2$ is superior to that of pure Pyrene. This difference can thus be ascribed in part to the lower oxidation onset potential of Pyr-$NH_2$, which helps to gain a higher amount of grafted Pyr-$NH_2$ molecules (10 wt % grafting).

The relation between the capacity per mass of the overall electrode and that of grafted Pyr-$NO_2$ is given in **Figure 5c**. A maximum 214 mAh $g^{-1}$ is obtained for Pyr-$NO_2$ at a grafting coverage of 5 wt %. However, as the grafting coverage increases, the capacity associated with the Pyr-$NO_2$ group decreases, which is presumably due to hindered mass transport within the porosity of the polymer. In order to utilize this *in situ* grafting strategy in a practical full cell, a high power density negative electrode is also under consideration. This is expected to replace the conventional Li, lithium titanate and pre-lithiated graphite negative electrodes.

The Nyquist plot in **Figure S8†** demonstrates that, even after 1000 cycles and a two-fold increase in the capacity of the OLC, there is only a slight rise in the charge transfer resistance in the case of the 20 wt % sample. By comparison, adding redox capacity to carbon by a conventional molecule mixture or covalent grafting is expected to lower the power density due to: (i) poor $e^-$ percolation through the carbon particle, since carbon particles are separated by the grafted molecules and by $sp^3$ defects of carbon as a result of grafting; and (ii) poor mass transport within the porosity of the molecular layer. By virtue of the present *in situ* grafting strategy, we are able to overcome the above constraints, as evidenced by the fact that the OLC architecture remains intact even after grafting, and that the kinetics improves. Indeed, as shown in **Figure 5d**, the power characteristics of the electrode with 20 wt % grafting is similar to that of the bare OLC electrode. Even at a current density of 10 A $g^{-1}$ (28 mA $cm^{-2}$), half of the capacity is retained. Interestingly, the decrease in capacity with respect to higher current density is similar to that of bare OLC (**Figure 5d**), demonstrating that *in situ* grafting as high as 20 wt % does not have any impact on the power characteristics. This result is in line with the Rct data from the EIS (**Figure S8†**). Self discharge characteristics are monitored by charging the cell to 4.4 V and followed by the voltage measurement for a relaxation for 10 hour and. Potential after 10 hour was 4.1 V in either case. As interestingly, the OLC/PPy-$NO_2$ lost merely 7.6 % of its capacity in 10 hour and while 11% loss observed in case of OLC. This was obvious as double layer capacitance shows more leakage current than that of redox capacity.

## CONCLUSION

In summary, we have developed a simple strategy for grafting Pyr-$NO_2$ molecules onto highly graphitized OLC via $\pi$-stacking, *in situ* reduction and a further polymerization of pyrene moieties at low potential. This approach facilitates the attainment of a high mass polymer loading, which results in an enhanced electrode redox capacity: when compared to Pyr-$NH_2$, Pyrene and even unreduced Pyr-$NO_2$, this demonstrates the key role of the redox activity coupled with the reduced $NO_2$ moiety, which is functionalized on the OLC backbone. DFT calculations demonstrate the charge transfer and binding enthalpy capabilities of Pyr-$NO_2$, which are significantly higher than those of the other two species, and which allow for improved p-stacking on the carbon surface. A 20 wt % grafting of Pyr-$NO_2$ enables a nearly two-fold increase in the capacity of the entire electrode from 20 mAh $g^{-1}_{electrode}$ to 38 mAh $g^{-1}_{electrode}$, while the average potential is increased by 200 mV. Under these conditions, the resulting grafted polymer shows a capacity of 110 mAh $g^{-1}$. Most importantly, the capacity retention is maintained for more than 4000 cycles without changing either the electrolyte or the Li counter electrode. The power characteristics of the bare OLC are also sustained, which further demonstrates the substantial advantages incurred by adopting the proposed controlled *in situ* grafting technique.

## Acknowledgements

BA, TB and JG acknowledge the financial assistance provided by the Agence Nationale de la Recherche (ANR) Project No. 13-PRGE-0011 and the French Research Network on Electrochemical Energy Storage (RS2E). YSAB and CE acknowledge EU H2020 Project ITN-MSCA-642742 "Enabling Excellence" for funding.

# Electronic Supporting Information

# Improved electro-grafting of nitropyrene onto onion-like carbon via *in situ* electrochemical reduction and polymerization: Tailoring redox energy density of the supercapacitor positive electrode

Bihag Anothumakkool[1], Yuman Sayed-Ahmad-Baraza[1], Chris Ewels[1], Pierre-Louis Taberna[2], Barbara Daffos[2], Patrice Simon[2], Thierry Brousse*[1] and Joel Gaubicher[1]*

1-Institut des Materiaux Jean Rouxel (IMN), University of Nantes, CNRS, 2, rue de Houssiniere-B.P. 32229-44322 Nantes cedex 3-France

2-Universite Paul Sabatier, Cirimat/Lcmie, 31062 Toulouse cedex 9 – France,

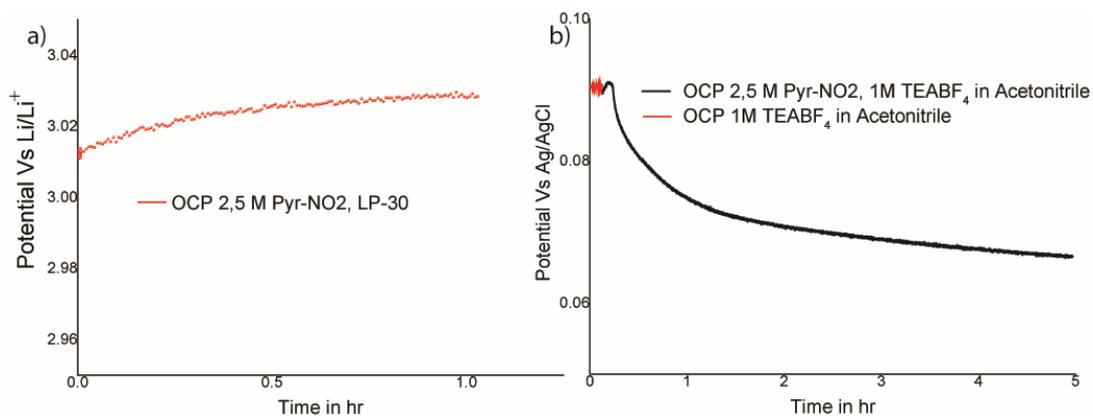

Figure S1: Open circuit potential (OCP) of YP-80 electrode in LP-30 and b) OCP of OLC electrode in 1M TEABF$_4$/Acetonitrile and further after the introduction of 2mM Pyr-NO2.

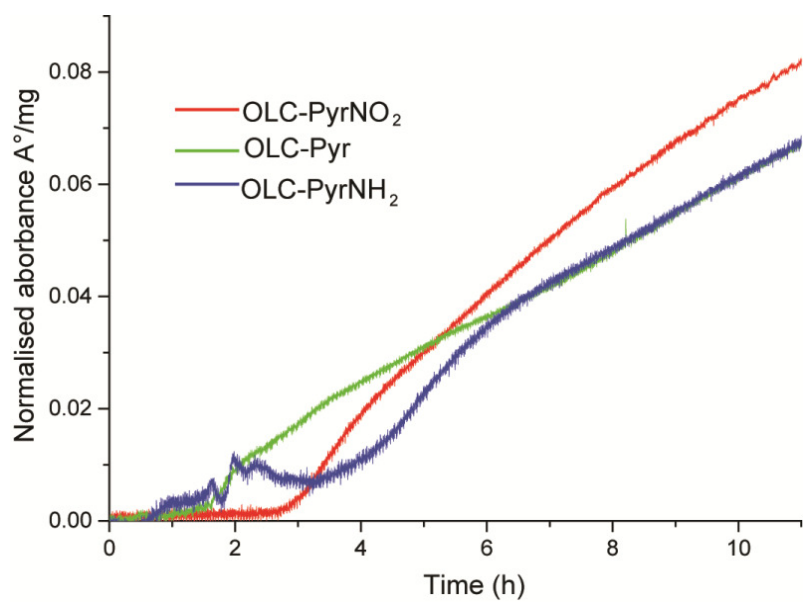

Figure S2: Adsorption of Pyrene and deriviatives onto carbon onion; in all cases the concentration was 0.25 mM in LP-30

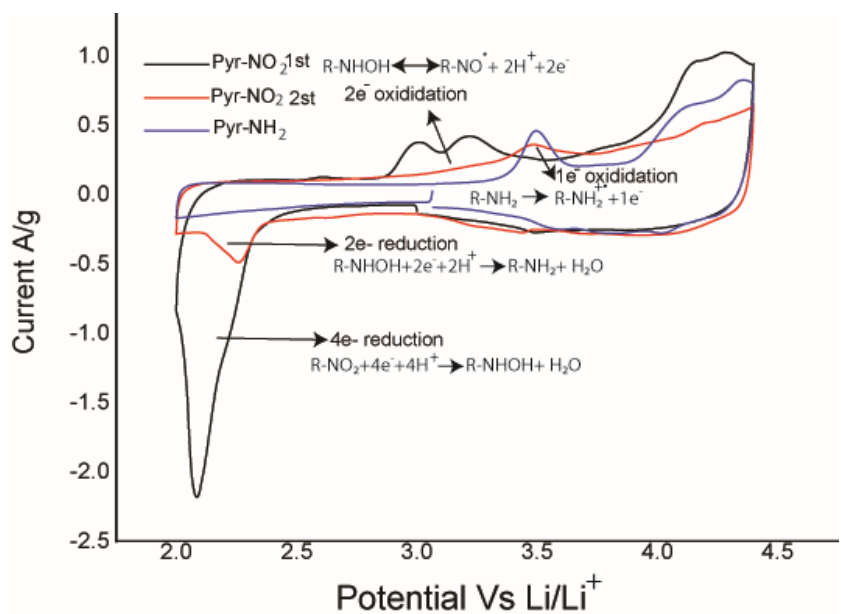

Figure S3: Initial CV profile at a scan rate of 5 mV/s in LP-30 containing 2 mM Pyr-NO$_2$ and Pyr-NH$_2$. Different electrochemical process and their corresponding peaks are indicated in the inset of figure.

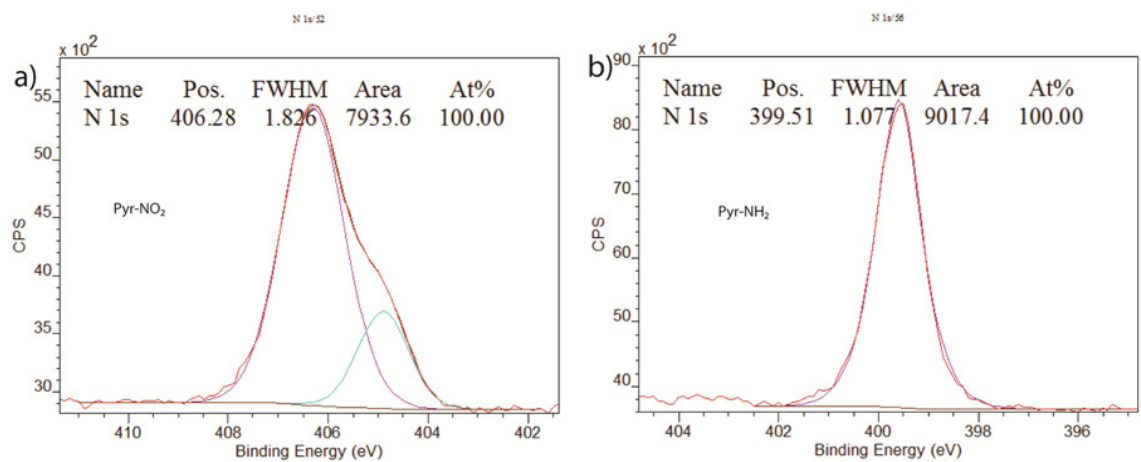

Figure S4: N1S XPS spectra of  a) 1-nitropyrene and b) 1-aminopyrene

Table S1. N1S core levl XPS details of OLC/PPyrNO2

| Peak name | Position | Full width Half maximum (FWHM) | % area |
|---|---|---|---|
| -NH$_2$ | 400.4 | 2.52 | 79.8 |
| -NHOH | 402.2 | 2.55 | 20.2 |

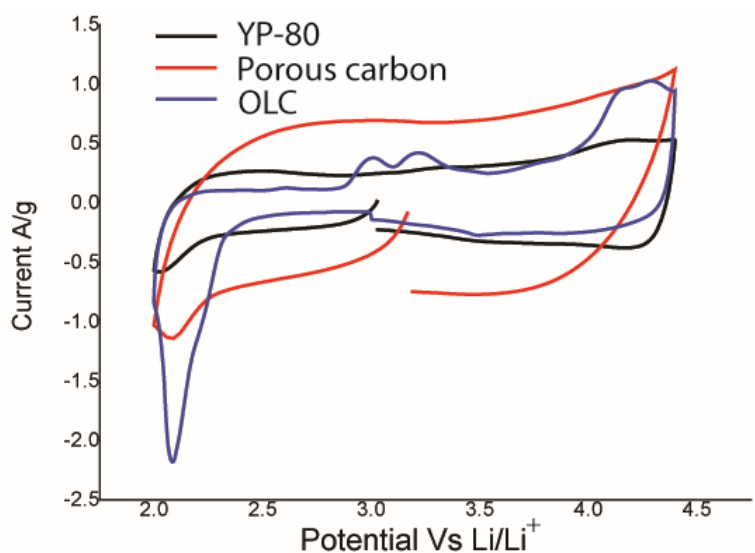

Figure S5: CV profile of OLC in comparison to two different commercial porous carbon electrode at a scan rate of 5 mV/s in LP-30 containing 2 mM Pyr-NO$_2$.

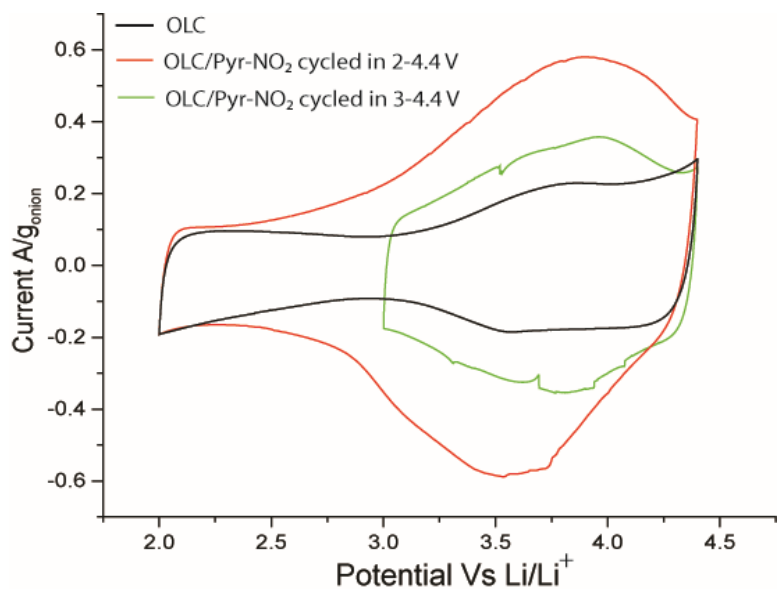

Figure S6: CV profile at a scan rate of 5 mV/s in LP-30 containing 2 mM Pyr-NO$_2$ derivative after electropolymerisation for 800 cycles in 2-4.4 V (red) and 3-4.4 V (green) window in comparison to bare carbon onion (black).

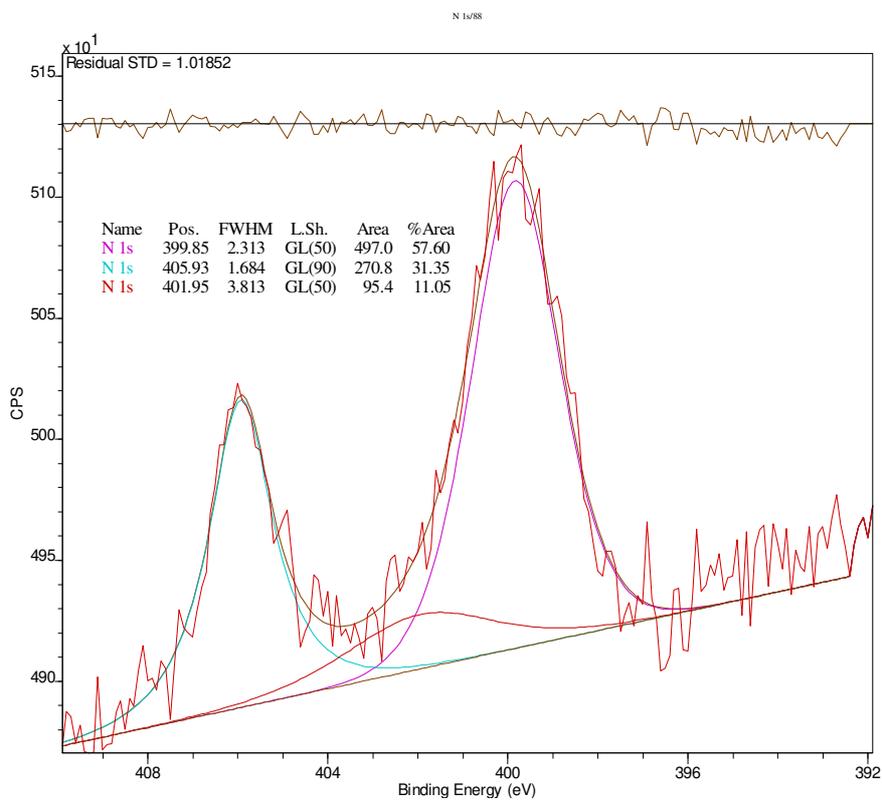

Figure S7: N1S core XPS spectra of PPyr-NO$_2$-coated OLC after cycling in a potential window of 2-4.4 V.

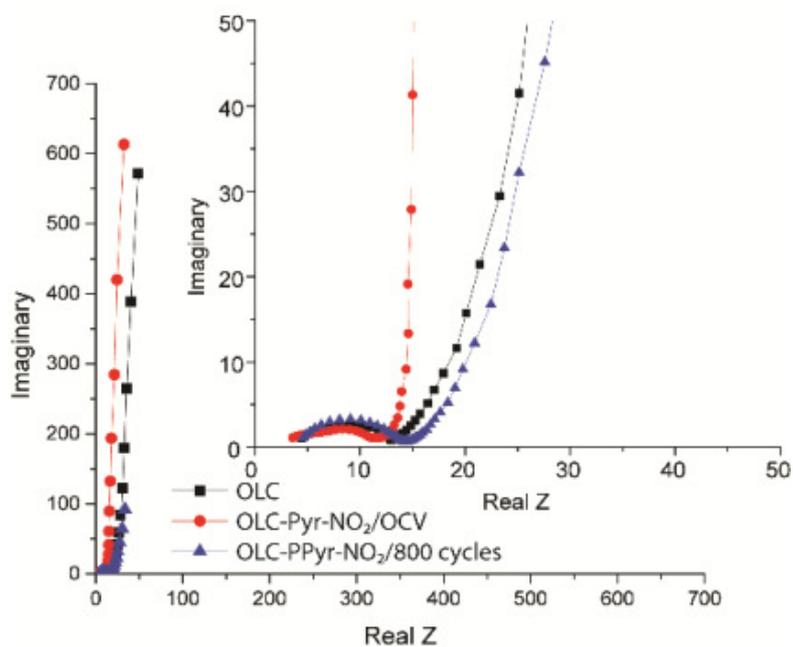

Figure S8: Nyqusit plot from the EIS spectra

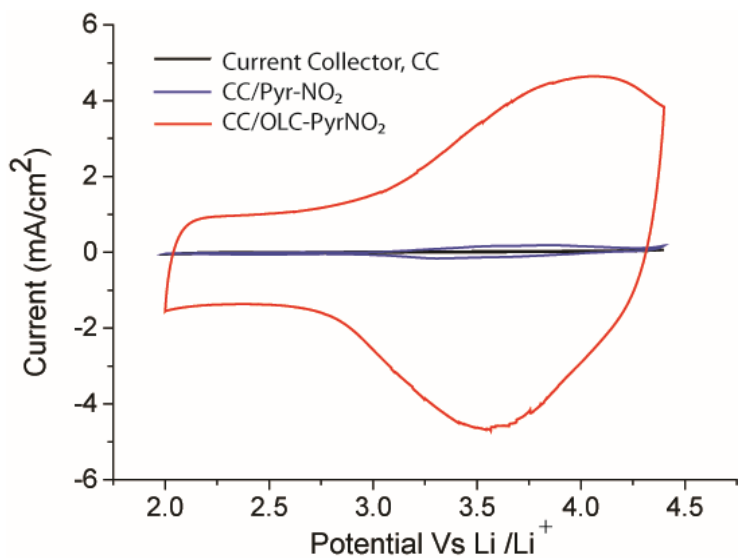

Figure S9: Areal current Vs voltage profile of OLC-PyrNO$_2$ in comparison to PyrNO$_2$ deposited onto current collector and pure current collector.

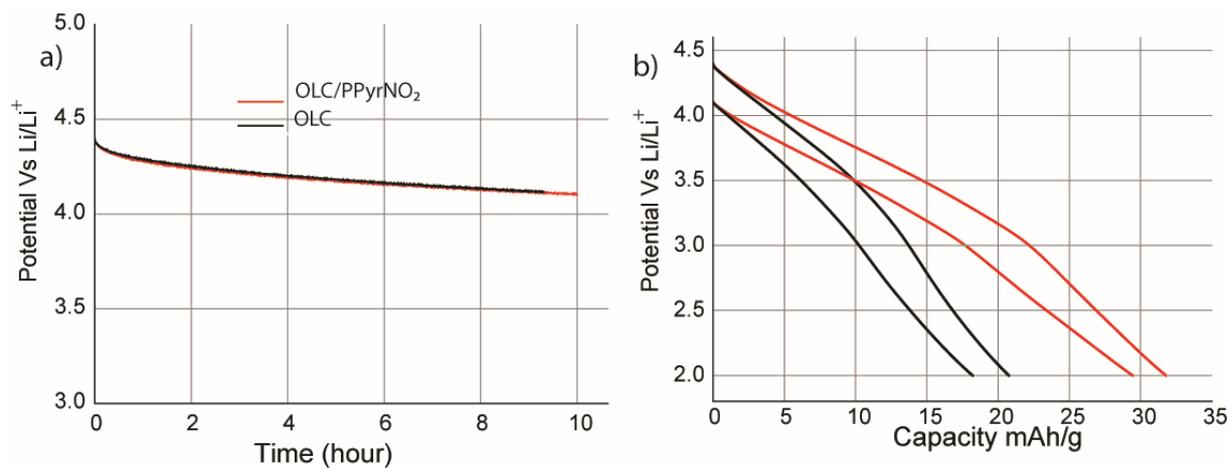

Figure S10: a) Plot of cell potential *vs.* relaxation time after charging to 4.4 V vs. Li$^+$/Li and b) change in the capacity due to the leakage loss during the relaxation time for 10 hour.